\def\ERLER{9912024}
\documentclass{article}		
\usepackage{latexsym}		  

\begin{document}
\begin{titlepage}

\title{{\Large\bf Dimensionally Democratic Calculus and Principles of
Polydimensional Physics}
\thanks{Financial Support for attending the conference provided
by Santa Clara Univ. }}
\author{{\Large\bf William M. Pezzaglia Jr.}
\thanks{wpezzag@clifford.org}
\\ Department of Physics
\\ Santa Clara University, CA }

\maketitle
\thispagestyle{empty}

\begin{abstract}
A solution to the 50 year old problem of a spinning particle in curved
space has been recently derived using an extension of Clifford calculus
in which each geometric element has its own coordinate.  
This leads us to propose that all the laws of physics should obey new
polydimensional metaprinciples, for which Clifford algebra is the
natural language of expression, just as tensors were for general relativity.
Specifically, phenomena and physical laws should be invariant under
local automorphism transformations which reshuffle the physical
geometry.  This leads to a new generalized unified basis for classical
mechanics, which includes string theory, membrane theory and the
hypergravity formulation of
Crawford[J. Math. Phys., {\bf 35}, 2701-2718 (1994)].
Most important is that the broad themes presented can be exploited
by nearly everyone in the field as a framework to generalize both the
Clifford calculus and multivector physics.

{Summary of presentation, submitted to the
{\it Proceedings of the 
5th International Conference on Clifford Algebras
and their Applications in Mathematical Physics,
Ixtapa-Zihuatanejo, MEXICO, June 27-July 4, 1999\/},
R. Ablamowicz and B. Fauser eds.}{}

\end{abstract}

\end{titlepage}

\pagestyle{myheadings}

\font\sixsy=cmsy6

\def\SWITCH#1#2{#1}       
\def\NOTE#1{\tt {#1}}
\def\NOTES#1{{\tt #1}}                
\def\HBAR{{\mathchar'26\mkern-9muh}}  
\def\vector#1{\vec{\bf {#1}}}         
\def\norm#1{\parallel{#1}\parallel}   
\def\LLL{{\cal L}}                    
\def\EQN#1{eq.\ (#1)}                 
\def\EB#1{{\bf e}_{#1}}
\def\EBU#1{{\bf e}^{#1}}
\def\EBH#1{{\bf \hat{e}}_{#1}}
\def\HALF{{1 \over 2}}
\def\HALFM{2^{-1}}
\def\FOURTH{{\scriptstyle 1 \over \scriptstyle 4}}    
\def\ODOT#1{\stackrel{\circ}{{#1}}}   
\def\ODOTS#1#2{\ODOT{#1}{\!}^{#2}}
\def\ODOTL#1#2{\ODOT{#1}{\!}_{#2}}
\def\VARO#1#2{\delta\!\!\ODOT{#1}{\!}^{#2}}

\def\COV#1{\nabla\!_{#1}}             
\def\PV#1#2{{\delta {#1} \over \delta {#2} }}
\def\PVE#1{\PV{\phantom{#1}}{#1}}       
\def\PD#1#2{{\partial {#1} \over \partial {#2}}}
\def\PDE#1{\PD{\phantom{#1}}{#1}}
\def\PP#1#2#3{{\partial^2 {#1} \over \partial {#2} \partial {#3}}}
\def\BD#1#2{{\left[\partial_{#1},\partial_{#2}\right]}}
\def\CD#1#2{\left[ \partial_{#1}, \partial_{#2} \right]}
\def\RRR{{\cal R}}
\def\POLYM{{\bf \cal M}}
\def\DDD{{\bf d}}
\def\TENSORL#1#2#3{#1_{#2}^{\phantom{#2}\!#3}}
\def\TENSORU#1#2#3{#1^{#2}_{\phantom{#2}\!#3}}
\section*{I. Introduction}
Taking an existing equation and generalizing it over quaternions
or Clifford numbers is certainly a way of doing new mathematics.
It is important however to understand that this seldom leads to
new physics (for example complexifying Newton's law of gravitation
is meaningless).  Reformulating existing physical
laws with a new mathematical language will \underbar{not} lead to
new principles nor new physics.  Only by generalizing principles
can we hope to do something new.  However, because Clifford
algebra\cite{Hestenes65}
encodes the structure of the underlying geometric space, we see
possible bigger patterns emerge.  Specifically in the description
of a spinning particle the equations of motion are invariant
under a non-dimensional preserving {\it polydimensional}\/
transformation which rotates between vector momentum and
bivector spin.  We therefore propose that `what is a vector' is
{\it Dimensionally Relative}\/ to the observer's frame, and that
the universe is fully {\it Polydimensionally Isotropic} in that
there is no absolute `direction' which we can assign `vector'
geometry over bivector, trivector, etc.

This forces us to propose a fully {\it Dimensionally Democratic}
Clifford calculus, in which each geometric element has its own
coordinate in a {\it Clifford manifold\/}\cite{Chisholm}.  In 
particular we show the utility of this concept in treating the
classical spinning particle in several scenarios.
A new action principle is proposed in which particles take
paths which minimize the sum of the linear distance traveled
combined with the bivector area swept out by the spin.
In curved space, the velocity of the variation is not the variation
of the velocity, leading to a new derivation of the Papapetrou
equations\cite{Papa} as autoparallels in the Clifford manifold.
This leads us to propose that the physical laws might be 
{\it Metomorphic Covariant\/} under general automorphism
transformations which reshuffle the geometry.

\section*{II. Relative Dimensionalism}
Most physicists tend to be absolute in their association of physical
quantities to geometric entities.  For example, mass is a scalar while
force a vector.  The introduction of Einstein's relativity however
caused a shift in dimensional interpretation in that `the world' is not
three-dimensional but a four-dimensional spacetime continuum.  In the
`old-fashion' three dimensional viewpoint, energy was a scalar, but in
the new 4D paradigm it is the fourth component of the momentum vector.
Certainly many physicists will share the opinion that the new 4D
viewpoint is right, and the old 3D view is incorrect.  Yet let us
consider for example the recent book by Baylis\cite{Baylis} in which
he has a complete treatment of electrodynamics and special
relativity using {\it paravectors}\/ (defined as the
Clifford aggregate of a scalar plus three-vector).  Is he wrong to
call time the scalar part of a paravector instead of calling it
the fourth component of a vector?

Consider also that the even subalgebra of the 16 element Clifford
algebra associated with 4D spacetime can be interpreted as the 
Clifford algebra of a 3D space.  Three of the planes of 4D are
reinterpreted as basis vectors in the 3D space, while the four-volume
is reinterpreted as a three-volume.  So if you grab a particular
geometric element, is it a vector or a plane?  We suggest that
there is no absolute right or wrong answer.
We postulate a new principle of {\bf relative dimensionalism}\/:
{\it that the geometric rank an observer assigns to an object is 
a function of the observer's frame of reference}\/ (or perhaps
state of conciousness).  There is no ``absolute'' dimension that
one can assign to a geometric object.  Further we consider
transformations which reshuffle
the basis geometry (e.g. vector line replaced  by bivector plane),
yet leave sets of physical laws invariant.  One
application provides a new treatment of the classical spinning 
particle, showing that the mechanical mass is enhanced by the
spin motion.

\subsection*{A.  Review of Special Relativity}
It is useful to see how paradigm shifts in the concept of the 
dimensional nature of space have impacted the formulation of
physical laws in the past for clues as how to proceed with newer
ideas.  The {\it prima facie} example is how things changed with the
introduction of special relativity.  Quantities such as time
and energy, that were previously defined as scalars (in a 
3D formulation) now are identified as fourth components
of vectors in four-dimensional Minkowski spacetime.  Let us consider
what is gained by using the higher dimensional concept.

\subsubsection*{1.  Unification of Phenomena}
The most obvious advantage of using vectors is that one can replace
a set of physical equations by a single vector equation.  Let us 
consider the application of four-vectors in electrodynamics.  The
3D scalar work-energy law and 3D vector force law,
$$\dot{{\cal E}} = e\, \vector{E} \cdot \vector{v}\ ,
\quad\quad \dot{\vector{P}} = e\, \left( \vector{E} + \vector{v}\times
\vector{B} \right)\ , \eqno(1ab)$$
can be combined into one single equation,
$$\dot{p}^\mu = \left({e \over mc}\right) p_\nu\, F^{\mu\nu}\ ,
\eqno(2)$$
using 4D vectors and tensors ($c$ is the speed of light and the 
dot represents differentiation with respect to time).
Certainly the adoption of the four-dimensional
viewpoint has notational economy, and provides insight that the
work-energy theorem (1a) is simpy the fourth aspect of the vector
force law (2).  However, philosophically one can ask if the 4D viewpoint
is any more correct than the 3D equations as they describe the
same phenomena.  Since special relativity
was originally formulated without the concept of Minkowski spacetime,
it is convenient, but apparently not necessary to
adopt the paradigm shift from 3D to 4D.
Hence we are being purposely dialectic in raising the question 
whether one can make an absolute statement about the dimensional
nature of a physical quantity such as time.  Can we state
(measure) that time is a part of a four-vector (as opposed to
a 3D scalar), or is this relative to whether one adopts a 3D or
4D world view, hence relative to the observer's dimensional 
frame of reference?

\subsubsection*{2.  Lorentz Transformations}
In classical physics the fundamental laws must be invariant under
rotational displacements because it is postulated that the universe
is {\it isotropic} (has no preferred direction).
When one formulates laws with vectors (which are inherently coordinate
system independent),
isotropy is `built in' without needing to separately impose the
condition.  Hence (Gibbs) vectors are a natural language to express
classical (3D) physical laws because they naturally encode isotropy.

Einstein further postulated the metaprinciple that motion was relative;
that there is no absolute preferred rest frame to the universe.  This
coupled with the postulate that the speed of light is the same for all
observers leads to the principle that the laws of physics must be invariant
under Lorentz transformations (which connect inertial frames of reference).
A more geometric interpretation is to see that Lorentz transformations
are just rotations in 4D spacetime, hence the principle of relative
motion is an extension of the metaprinciple of isotoropy to four-space.

Lorentz transformations, expressed in four-space, preserve the rank of
geometry (rotates a four-vector into another four-vector).  In contrast,
Baylis\cite{Baylis} would write the Lorentz boost (in the 
$z$ direction with velocity $v$) of the
{\it momentum paravector}\/ in 3D Clifford algebra as,
$${\bf 1}{\cal E}^\prime/c + \vector{P}^\prime = 
{\cal R}\left({\bf 1}{\cal E}/c + \vector{P} \right) {\cal R}^\dagger \ ,
\eqno(3)$$
where $c$ is the speed of light, and the transformation operator:
${\cal R} = \exp(-\EBH{3} \beta /2)$, where
$\beta$ the rapidity related to the
velocity: $\tanh(\beta) = v/c$ and $\EBH{3}$ is the unit vector in the
$z$ direction.  As a consequence, in this 3D perspective, what is
pure scalar (e.g. ${\cal E}=$energy)
to one observer is part scalar, part vector to another
observer.  Lorentz transformations, in this 3D viewpoint, are
NOT dimensional preserving.  We choose to classify such transformations
as {\it geometamorphic\/} or `polydimensional'.

\subsubsection*{3.  Invariant Moduli}
In 3D space the length (magnitude) of a vector (e.g. electric field
or momentum) is invariant under rotations.  Under Lorentz transformations
(4D rotations), the modulus of the four-vector is invariant,
$$\norm{{\bf p}}^2\equiv p_\mu p^\mu = 
{\cal E}^2/c^2 -\norm{\vector{P}}^2\ . \eqno(4)$$
Reinterpreted with a 3D viewpoint, the invariant quantity of the
Lorentz transformation (3) is the
difference between the square of the scalar energy minus the magnitude
of the 3D momentum vector.
Neither the modulus of the 3D scalar, nor 3D vector is
independently invariant under these transformations.

The modulus of the momentum four-vector (4a) is defined to be the
{\it rest mass}\/ of the particle: $m_0 \equiv c^{-1} \norm{\bf p}$.
When in motion, the mechanical mass of the particle (e.g. in 
definition of momentum: $p=mv$) increases by its kinetic energy
content.  This is described by the {\it Lorentz dilation
factor\/} $\gamma$,
$$m \equiv \gamma\,m_0 \ , \eqno(5a)$$
$$\gamma \equiv \cosh{\beta}=
\left(1 - {v^2 \over c^2}\right)^{-\HALF} =
\sqrt{1 + \left({\norm{\vector{P}} \over m_0 c}\right)^2}\ . \eqno(5b)$$

\subsection*{B.  Automorphism Invariance}
Transformations that preserve some physical symmetry often lead to
a conservation law of physics.  For example, displacements of the
origin leave Newton's laws unchanged, leading to a derivation of
the conservation of linear momentum.  Central force problems (e.g.
gravitational field around a spherical star) have rotational
invariance, leading to conservation of angular momentum for orbits.
When formulating physics with Clifford algebra, we should ask just
what new symmetries are inherently encoded in the structure of the
algebra, and what (if any) new physical laws they may imply.

\subsubsection*{1.  Matrix Representation Invariance}
Physicists usually first encounter Clifford algebras in quantum mechanics
in the form of Pauli, Majorana and Dirac `spin' matrices.  The matrix  
representation for example of the four generators $\gamma^\mu$ of the
Majorana algebra is arbitrary.  Hence it is obvious that the physics
must be invariant under a change of the matrix representation of the
algebra.  A change in representation
can be reinterpreted as a global rotation of the spin
space basis spinors.  Requiring `spin space isotropy' (no preferred
direction in spin space) leads to the physical principle of conservation
of quantum spin. 

\subsubsection*{2.  Algebra Automorphisms}
Its possible however to avoid talking about the matrix representation 
entirely.  The more general concept is an algebra automorphism, which
is a transformation of the basis generators $\gamma_\mu$ of the
algebra which preserves the Clifford structure,
$$\{ \gamma_\mu , \gamma_\nu \} = 2\,g_{\mu\nu}\ , \eqno(6) $$
where $g_{\mu\nu}$ is the spacetime metric.  For example, 
consider the following orthogonal
transformation on any element ${\cal Q}$ of the Clifford algebra,
$${\cal Q}^{\prime} = {\cal R}\, {\cal Q}\, {\cal R}^{-1}\ , \eqno(7)$$
$${\cal R}(\phi^\mu)\equiv \exp(-\gamma_\mu\,\phi^\mu /2)\ ,
\quad\ \mu=1,2,3,4.  \eqno(8)$$
Proposing local covariance of the Dirac equation under such
automorphism transformations is one path to gauge gravity and 
grand unified theory\cite{Crawford}.

\subsubsection*{3.  Polydimensional Isotropy}
If the elements $\gamma_\mu$ are interpreted geometrically as basis
vectors, then (8) reshuffles geometry.  For example, when $\phi^4=
\pi /2$, \EQN{7} causes the permutation,
$$\gamma_j \Longleftrightarrow
\gamma_4 \gamma_j\ , \ \ \ j = 1,2,3, \eqno(9a)$$
$$\gamma_1 \gamma_2 \gamma_3 \Longleftrightarrow
\gamma_4 \gamma_1 \gamma_2 \gamma_3 \ , \eqno(9b)$$
which exchanges three of the vectors with their associated timelike
bivectors.  What is a 1D vector in one ``reference frame'' is hence a
2D plane in another.  The transformation (8) thus ``rotates'' vectors
into planes.  Another example would be the transformation generated
by (7) for: ${\cal R} = \exp(\hat{\epsilon}\, \phi /2 )\/$, where
$\hat{\epsilon} \equiv \gamma_1 \gamma_2 \gamma_3 \gamma_4$.
When $\phi=\pi/2$ this causes the duality transformation,
$$\gamma_\mu \Longleftrightarrow \hat{\epsilon}\,\gamma_j\ ,
\ \ \ \mu = 1,2,3,4, \eqno(10)$$
exchanging vectors for dual trivectors (the rest of the
algebra is unchanged).

If we feel that Clifford algebra is 
the natural description of a geometric space, then we must ask
whether these algebra automorphisms have physical interpretation.
We suggest that there is a `higher octave' to the metaprinciple of
the isotropy of space.  We propose the
{\bf principle of polydimensional isotropy}: that {\it there is 
no absolute or preferred direction in the universe to which one can
assign the geometry of a vector}\/.  Just as which direction you choose
to call the z-axis is arbitrary, it is also arbitrary just which 
geometric element you call the basis vector in the z-direction.
Another observer may make an entirely different selection.
This suggests perhaps that we should require the laws of physics to be 
invariant under such automorphic transformations (which will be
discussed in more detail in Section IV).

\subsection*{C.  Polydimensional Formulation of Physics}
If we embrace the new principle that space is polydimensionally 
isotropic, then we should consider if transformations that reshuffle the
basis geometry leave certain sets of physical laws invariant.
Our development will attempt to parallel that which happened in
the transition from classical 3D physics to 4D special relativity.

\subsubsection*{1.  Unification with Clifford Algebra}
Just as four-vectors allowed us to unify two equations into
one, the language of Clifford algebra allows for further notational
economy.  Consider that a classically spinning point charged particle
obeys the torque equation of motion\cite{Rohrlich},
$$\dot{S}^{\mu\beta} = \left( {e \over mc} \right) \left(
F^\mu_{\,\ \nu}\> S^{\nu\beta} - F^\beta_{\,\ \nu} \>
S^{\nu\mu} \right) =\left( {e \over mc} \right) 
\left( {\bf F} \otimes {\bf S}\right)^{\mu\beta}\ . \eqno(11)$$
This and \EQN{2} can be written in the single statement,
$$\dot{\POLYM} = \left( { e \over 2mc } \right) \left[
{\bf F},\POLYM \right]\ , \eqno(12a) $$
where ${\bf F}\equiv{1 \over 2} F^{\mu\nu}\,\EBH{\mu}\wedge\EBH{\nu}$
is the electromagnetic field bivector and $\EBH{\mu}$ are the basis
vectors of the geometric space which obey the same Clifford algebra
rules \EQN{6}.  The {\it momentum
polyvector} is defined as the multivector sum of the
vector linear momentum and the bivector spin momentum,
$$\POLYM \equiv p^\mu \EBH{\mu} + {1 \over 2 \lambda}
S^{\mu\nu} \EBH{\mu} \wedge \EBH{\nu}\ , \eqno(12b)$$
where $\lambda$ is some fundamental length scale constant (to be
interpreted in the next section) that allows us to add the quantities
with different units in analogy to the use of $c$ in \EQN{3}.
The ability to add different ranked (dimensional) geometries
is the notational advantage of Clifford geometric algebra over
standard tensors.  Mathematically, (12a) allows one to
{\it simultaneously} obtain solutions to both equations (2) and (11):
$\POLYM (\tau) = \RRR\>\POLYM(0)\>{\cal R}^{-1}$, where the 
rotation operator,
$$\RRR(\tau) = \exp\left( {e \over 4mc}\, \EBH{\mu}\wedge \EBH{\nu}
\int^\tau d\tau^\prime 
F^{\mu\nu}\left[ x(\tau^\prime)\right] \right) \ , \eqno(13)$$
involves a path (history) dependent integral, hence the solution is formal.

\subsubsection*{2.  Polydimensional Invariance}
Equation (12a) is manifestly covariant under automorphism transformations.
Specifically, the set of equations (2) and (11) are invariant under the
automorphism transformations generated by (8).  For example,
$\phi^4=\pi /2$ in (8) causes a trading between momentum and
mass moment of the spin tensor,
$$\lambda \> p_j \Longleftrightarrow  S_{4j}\ . \eqno(14)$$
It is not at all clear what physical interpretation to ascribe to the
two frames of reference.  A radical assertion of the {\it principle
of relative dimensionalism} would be to propose that
what is a vector to one observer is a bivector to another, and that 
they would partition the polymomentum (12b) into momentum and spin
portions differently.  {\it What is spin to one would be momentum to
the other.}

Just as rotational invariance led to conservation of angular momentum,
we might ask just what is the conserved quantity associated with this
new symmetry transformation.  This will be addressed in 
Section III.B.3 below.
  
\subsubsection*{3.  The Quadratic Form of a Polyvector}
We define the modulus of the polyvector \EQN{12b} to be the square
root of the scalar part of the square of the polyvector,
$$\norm{ \POLYM}^2 \equiv \> p_\mu\,p^\mu
-{1 \over 2 \lambda^2} S_{\mu\nu}\,S^{\mu\nu}\ . \eqno(15)$$
This quadratic form is invariant under the rotation of vectors into
bivectors generated by (8).  In the $(---+)$ metric signature, we 
define the modulus of the momentum polyvector to be the {\it bare mass\/}:
$m_0 \equiv c^{-1} \norm{{\POLYM}}$.
This implies that the mechanical mass (modulus of the momentum)
is NOT invariant under these transformations, but has been
enhanced by the spin energy content,
$$m \equiv c^{-1} \norm{{\bf p}} = m_0  \sqrt{1 + 
{S^{\mu\nu} S_{\mu\nu} \over 
2\left( m_0 c \lambda \right)^2}}\ , \eqno(16a)$$
in analogy to (5ab).  What we have described in (16a), by simple
geometric construction, is a familiar result, laboriously obtained by
Dixon\cite{Dixon70} in the mechanical analysis of extended spinning
bodies.  Expanding (16a) non-relativistically,
$$m c^2 \simeq m_0 c^2 + \left({ \vector{S}^2
\over 2 m_0 \lambda^2 } \right) + \dots \ , \eqno(16b)$$
where $\vector{S}^2 \equiv \left( S_{12} \right)^2 +
\left( S_{23} \right)^2 +\left( S_{31} \right)^2$,
one sees that $\lambda$ is consistent with the {\it radius of
gyration}\/ of a classical extended particle such that its moment of 
inertia is $\simeq m_0 \lambda^2$.  Hence the correction
to the mass is due to the rotational kinetic energy.

\section*{III.  Dimensional Democracy}
In quaternionic analysis, a coordinate is given to 
each of the four elements.  Reinterpreted as a Clifford algebra,
it would be as if one has given a coordinate to each of the two
vector directions, one to the plane, and one to the scalar.
We now propose that each geometric element of the Clifford algebra
democratically has its own conjugate coordinate.  Further {\it the
physical laws should be multivectorial, with each geometric
component meaningful\/}.  For example, our polymomenta (12)
gives the (vector) linear momenta and
(bivector) spin momenta equal importance, both contributing the
the modulus of the vector.  This suggests a
generalized action principle that particles take the paths
which minimize the sum of the linear distance traveled combined
with the bivector area swept out.  This simple geometric idea gives
a new derivation of the spin enhanced mass described by the Dixon
equation (15), as well as proposals for new quantum equations.

\subsection*{A.  Review of Classical Relativistic Mechanics}
In ancient times, Heron of Alexandria showed that light reflecting off
a mirror would take the path of least distance between two
endpoints.  The generalized concept is that classical
particles will follow paths of least spacetime distance between
endpoints, even when the space is curved by gravity.

\subsubsection*{1. Time Contributes to Distance}
The measure of distance between two points in flat spacetime is,
$$c^2 d\tau^2 \equiv c^2 dt^2 - \left(dx^2 + dy^2 + dz^2
\right)= dx^\alpha dx^\beta \, g_{\alpha\beta} \ , \eqno(17a)$$
where affine parameter $\tau$ is commonly called the {\it proper time}.
If we adopt the 3D viewpoint, we are combining (in quadrature)
the `scalar' time displacement with the `vector' path displacement,
utilizing a fundamental constant $c$ (the speed of light)
to combine the quantities which have different units.  The metric tensor
$g_{\alpha\beta}$ in flat space is diagonal with elements $(-1,-1,-1,+1)$
such that the fourth time component has the opposite signature of the
spatial parts.  To generalize for curved space,
$g_{\alpha\beta}(x^\sigma)$  becomes a function of spacetime position.

Dividing (17a) by $dt^2$ recovers the {\it Lorentz dilation
factor}  \EQN{5b} in terms of the nonrelativistic velocity,
$$\gamma \equiv {dt \over d\tau} = \left(1 - {v^2 \over c^2}
\right)^{-1/2}\ . \eqno(17b)$$

\subsubsection*{2.  Euler-Lagrange Equations of Motion}
In simplest form, to obtain the equations of motion, one chooses the
special path between fixed endpoints for which the
{\it action integral\/} [which is based upon the quadratic
form \EQN{17a}],
$${\cal A} \equiv \int m_0 c\ d\tau = 
\int {\cal L} d\tau = 
\int m_0 c \sqrt{u^\alpha\,u^\beta
\, g_{\alpha\beta}(x^\sigma) } \ d\tau\ , \eqno(18)$$
is an {\it extremum\/}.
The integrand ${\cal L}(\tau,x^\alpha,\dot{x}^\alpha)$ is called
the {\it Lagrangian}, which is generally a function of the
coordinates and their velocities relative to the proper time: 
$u^\alpha \equiv \dot{x}^\alpha = dx^\alpha / d\tau$,
where $x^4\equiv ct$, hence $u^4=\dot{x}^4 = c\gamma$.

Each coordinate
$x^\alpha$ has a canonically conjugate momentum $p_\alpha$ defined,
$$p_\mu \equiv {\delta {\cal L} \over \delta u^\mu} = m_0 u_\mu
=m_0 \dot{x}_\mu\ . \eqno(19a) $$
For our relativistic Lagrangian (18) these obey \EQN{4}.
When reparameterized in terms of the more familiar
observer's time $t=x^4/c$,
$$\hbox{\tt Momentum:}\quad P_j \equiv \PV{\cal L\ }{\dot{x}^j}= 
m_0 \dot{x}_j = m\,v_j\ ,\quad j=1,2,3, \eqno(19b)$$
$$\hbox{\tt Energy:}\quad{\cal E}\equiv c\PV{\cal L\ }{\dot{x}^4}=
c\,m_0\,\dot{x}_4 = m\,c^2\ ,\quad\quad\ \quad \eqno(19c)$$
it is easy to show that the 3D part of the
momentum $P_j=m v_j$ has mass $m$ which is enhanced by the
energy content according to (5ab).

Applying Hamilton's {\it Principle of Least Action}, one
considers the total
variation of the Lagrangian with respect to a variation
in path (and velocity),
$$\delta{\cal L} = \PV{\cal L}{x^\alpha}\>\delta x^\alpha +
\PV{\cal L}{\dot{x}^\alpha}\>\delta\dot{x}^\alpha\ . \eqno(20a)$$
To get the equations of motion as that part proportional to a variation
in the path only, the last term involving the variation of the 
velocity is integrated by parts,
$$\PV{\cal L}{\dot{x}^\alpha}\delta\dot{x}^\alpha\;\equiv
\;p_\alpha \delta\dot{x}^\alpha\;=
\;{d \over d\tau} \left(p_\alpha \delta x^\alpha \right) -
\dot{p}_\alpha \delta{x}^\alpha
+p_\alpha \left(\delta\dot{x}^\alpha - {d \over d\tau}
\delta x^\alpha \right)\ . \eqno(20b)$$
The total derivative term does not contribute if the variation in path has
fixed endpoints.  Substituting into (20a) and setting equal to zero,
one obtains the generalized Euler-Lagrange
equations of motion\cite{Fiziev},
$$\left(\dot{p}_\alpha - \PV{\cal L}{x^\alpha}\right)\delta x^\alpha=
p_\beta \left(\delta\dot{x}^\beta - {d \ \over d\tau} 
\delta x^\beta \right) \ . \eqno(20c)$$
In most elementary mechanics texts the terms on the right are argued
to vanish because it is assumed the variation of the velocity is the
same as the velocity of the variation.  However when coordinates are
path dependent (non-holonomic), $\delta \DDD \not=\DDD \delta$ because
derivatives will not commute\cite{Fiziev, Kleinert}.
For example, in a rotating coordinate system,
$$\left(\delta\dot{x}^\beta - {d \over d\tau} \delta x^\beta
\right) = \delta x^\alpha\,\omega_\alpha^{\ \beta} \ . \eqno(20d)$$
This would introduce an $\vector{\omega}\times\vector{P}$ 
pseudoforce term
on the right side of the equation of motion \EQN{20c}.

\subsubsection*{3.  Rotating Coordinate Systems}
The principle of isotropy states that there is no preferred direction
in spacetime.  Hence the laws of physics must be invariant under
local Lorentz transformations (which include spatial rotations).
Hence we can invent a new ``body frame'' coordinate system which 
has time dependent basis vectors $\EB{\mu}(\tau)$ that are
related to the fixed ``lab frame'' basis $\EBH{\mu}$ by a time
dependent orthogonal transformation,
$$\EB{\mu}(\tau) = {\cal R}\,\EBH{\mu}\,{\cal R}^{-1} \ , \eqno(21a)$$
$${\cal R}(\tau) = \exp\left(- \FOURTH\EBH{\mu\nu}
\,\Theta^{\mu\nu}(\tau) \right) \ . \eqno(21b)$$
The cartesian angular displacement coordinates $\Theta^{\mu\nu}$
uniquely describe the orientation state of the body frame at the
particular time.  However, the angular velocity bivector
$\underline{\omega}$ of the frame is NOT given by the time derivatives
of these coordinates, rather its defined\cite{Hestenes90},
$$\underline{\omega} \equiv (1/2) \omega^{\mu\nu}\,\EBH{\mu\nu}
\equiv -2\,{\cal R}^{-1}\dot{\cal R}\quad \not=
\quad (1/2)\,\dot{\Theta}^{\mu\nu}\,\EBH{\mu\nu} \ . \eqno(22)$$
The difficulty is that rotations (Lorentz transformations) do not 
commute, hence the final state of the body frame is a function of 
path history.  This was also the case in the
electrodynamic problem presented earlier in \EQN{13}.

One can invent some new {\it quasi-coordinates}: $\theta^{\mu\nu}$,
for which the angular velocity IS given by their
time derivative,
$\omega^{\mu\nu} \equiv \dot{\theta}^{\mu\nu}$,
(see Greenwood\cite{Greenwood}).
Unfortunately, these new coordinates are non-integrable
(path dependent) and hence {\it non-holonomic} such that
$(\delta \DDD - \DDD \delta )\theta^{\mu\nu} \not=0$.  The advantage of
resorting to this complexity is that the Lagrangian and (generalized)
Euler-Lagrange equations have the same form in both the body
frame and lab frame\cite{Fiziev}.  Further, we can show by the
chain rule,
$$\RRR\underline{\omega}=-2\dot{\RRR} = -2 \left(
{1 \over 2} \dot{\theta}^{\mu\nu} \PD{\RRR}{\theta^{\mu\nu}}
\right) \ , \eqno(23a)$$
that the tangent bivectors are given by the derivatives of the
rotation operator,
$$2 \PD{\RRR\ }{\theta^{\mu\nu}} = - \EB{\mu\nu}{\RRR}
= -{\RRR}\EBH{\mu\nu} \ . \eqno(23b)$$
The differential and variation of the rotation operator are hence,
$$\DDD {\cal R} = -\HALF {\cal R}\, d\theta^{\mu\nu}
\,\EBH{\mu\nu} \ ,
\quad\quad \delta \RRR = -\HALF \RRR\,\delta\theta^{\mu\nu}
\,\EBH{\mu\nu} \ . \eqno(24ab)$$

We assume that since $\RRR$ defines the state of the body 
{\it independent of the particular coordinate parametrization},
that $\delta(\DDD\RRR) = \DDD(\delta{\RRR})$.  Explicitly taking
the variation $\delta$ of \EQN{24a}, and setting it equal to the
differential $\DDD$ of \EQN{24b} we obtain,
$$\left(\delta\DDD - \DDD\delta \right) \HALF \theta^{\mu\nu}
\EBH{\mu\nu} = \RRR^{-1} \left(\DDD\RRR\,\delta\underline{\theta}
-\delta\RRR\,\DDD\underline{\theta} \right)
=\HALF \left[ \DDD\underline{\theta},
\delta\underline{\theta}\right] \ . \eqno(25a)$$
In component form, we see that for rotations, the variation of the
angular velocity bivector is \underbar{not} the velocity of the
angular variation bivector,
$$\left(\delta\omega^{\mu\nu} 
-{d \delta \theta^{\mu\nu} \over d\tau\ \ }\right)
= \left(\underline{\omega}\otimes \underline{\delta \theta}
\right)^{\mu\nu}=\delta^{\mu\nu}_{\alpha\beta}
 \>\omega^\alpha_{\ \sigma}\,\delta\theta^{\sigma\beta}
\ . \eqno(25b)$$

\subsection*{B.  Polydimensional Mechanics}
If we fully embrace the concept of {\it relative dimensionalism},
then we must recognize that what one observer labels as a `point'
in spacetime with vector coordinates $(x,y,z,t)$ may be seen as an
entirely different geometric object by another.  This suggests that
perhaps we should formulate physics in a way which
is completely {\it dimensionally democratic} in
that all ranks of geometry are equally represented.

\subsubsection*{1.  The Clifford Manifold}
We propose 
therefore that `the world' is not the usual four-dimensional manifold,
but instead a fully {\it polydimensional continuum}, made of points,
lines, planes, etc.  Each event $\Sigma$ is a geometric point in a
{\it Clifford manifold}\/\cite{Chisholm},
which has a coordinate $q^A$ associated with each
basis element ${\bf E}_A$ (vector, bivector, trivector, etc.).
Our definition of the ``Clifford Manifold'' is hence broader than
the original proposal by Chisholm and Farwell\cite{Chisholm} in that
we have been ``dimensionally democratic'' in giving a coordinate
to each geometric degree of freedom.
The {\it pandimensional differential} in the manifold would be,
$$d\Sigma \equiv {\bf E}_A dq^A=\EB{\mu}dx^\mu +
{1 \over 2 \lambda} \EB{\alpha}\wedge\EB{\beta}\>da^{\alpha\beta}
+{1 \over 6 \lambda^2} \EB{\alpha}\wedge\EB{\beta}\wedge\EB{\sigma}
\>dV^{\alpha\beta\sigma} +\dots\>, \eqno(26a)$$
where in Clifford algebra it is perfectly valid to add vectors 
to planes and volumes (parameterized by the antisymmetric tensor
coordinates $dx^\mu,
da^{\alpha\beta}, dV^{\alpha\beta\sigma}$ respectively).

In analogy to (15), we propose that the quadratic form of the 
Clifford manifold would be the scalar part of the square of (26a),
$$d\kappa^2= \norm{d\Sigma}^2 \equiv dx^\mu dx_\mu +
{1 \over 2 \lambda^2} da^{\alpha\beta} da_{\beta\alpha}
+ {1 \over 6 \lambda^4}dV^{\alpha\beta\sigma}
dV_{\sigma\beta\alpha} +\dots\ . \eqno(26b)$$
The fundamental length constant $\lambda$ must be introduced in
\EQN{26ab} in order to add the bivector `area' coordinate contribution
to the vector `linear' one.  In analogy to (17a)
this new quadratic form suggests we define a new 
affine parameter $d\kappa=\norm{d\Sigma}$ which we will use to
parameterize our polydimensional equations of motion.

\subsubsection*{2.  New Classical Action Principle}
Classical mechanics assumes points which
trace out linear paths.  The equations of motion are based upon minimizing
the distance of the path.  String theory introduces one-dimensional
objects which trace out areas, and the equations of motion are analogously
based upon minimizing the total area.  Membrane theory proposes
two-dimensional objects which trace out (three-dimensional)
volumes to be minimized.  Our new action principle is
that we should add all of these
contributions together, and treat particles as polygeometric objects
which trace out polydimensional paths with (26b) the quantity to be
minimized.

Using only the vector and bivector contributions of (26b) the
Lagrangian that is analogous to (18) would be,
$${\cal L}(x^\mu,\ODOTS{x}{\mu},a^{\alpha\beta},
\ODOTS{a}{\alpha\beta} )= m_0 c 
\sqrt{\ODOTS{x}{\mu} \ODOTS{x}{\nu}\,g_{\mu\nu}-{1 \over 2\lambda^2}
\ODOTS{a}{\alpha\beta}\ODOTS{a}{\mu\nu}\,g_{\alpha\mu}\,g_{\beta\nu}
}\ , \eqno(27)$$
where the open dot denotes differentiation with respect to the new
affine parameter $d\kappa$ (whereas the small dot with
respect to the proper time $d\tau$),
$$\ODOT{Q} \equiv { dQ \over d\kappa} = \dot{Q} 
{d\tau \over d\kappa}\ . \eqno(28)$$
The relationship of the new affine parameter to the proper time is 
is a new {\it spin dilation factor} analogous to the Lorentz dilation
factor (17b).  Dividing (26b) by $d\tau$ or $d\kappa$, and noting
$d\tau^2 \equiv dx^\mu \, dx_\mu$, this spin dilation factor is,
$${d\tau \over d\kappa}\equiv \left( 1 - {\dot{a}^{\mu\nu}
\, \dot{a}_{\mu\nu} \over 2 c^2 \lambda^2 } \right)^{-1/2}
= \sqrt{1 + {\ODOTS{a}{\mu\nu}\ODOTL{a}{\mu\nu}
\over 2 c^2 \lambda^2  }} \ . \eqno(29)$$
Note this implies the magnitude of the bivector velocity with respect
to the proper time (proportional to spin angular velocity) is
bounded by $\lambda\,c$, just as linear velocity cannot exceed $c$.

\subsubsection*{3.  Canonical Momenta}
In analogy to \EQN{19a} we interpret the spin to be the
canonical momenta conjugate to the bivector coordinate,
$$\hbox{\tt \quad\quad Spin:}\quad
S_{\mu\nu} \equiv \lambda^2 \PV{\cal L}{\ODOTS{a}{\mu\nu}} =
m_0 \ODOTL{a}{\mu\nu} = m\,\dot{a}_{\mu\nu}\ ,\quad \eqno(30a)$$
$$\hbox{\tt \quad Momentum:}
\quad p_{\mu} \equiv \PV{\cal L}{\ODOTS{x}{\mu}} =
m_0 \ODOTL{x}{\mu} = m\,\dot{x}_{\mu}\ ,
\quad\quad\quad\ \quad \eqno(30b)$$
$$\hbox{\tt Dynamic Mass:}\quad
m\equiv m_0 \ODOT{\tau}=m_0\,{d\tau \over d\kappa}
\ .\quad\quad\quad\quad\quad\quad\quad\quad \eqno(30c)$$
For our Lagrangian (27), these satisfy the Dixon
equation (15).  When these momenta are reparameterized in terms of
the more familiar proper time, they have spin enhanced mass
defined by (30c), which is equivalent to \EQN{16a}.

Our Lagrangian (27) is invariant under
the polydimensional coordinate
rotation (between vectors and bivectors), generated by four the
arbitrary parameters $\delta \phi^\alpha$ of the automorphism
transformation analogous to \EQN{8},
$$\delta x^\alpha = \lambda^{-1} 
\, a_{\ \mu}^{\alpha}\, \delta\phi^\mu\ , \eqno(31a)$$
$$\delta a^{\mu\nu} = x^\mu \,\delta\phi^\nu 
-x^\nu\,\delta\phi^\mu \ . \eqno(31b)$$
Noether's theorem associates with this symmetry transformation a 
new set of constants of motion,
$$Q_\mu =\PV{\cal L\ }{\ODOTS{x}{\alpha}} \PV{x^\alpha}{\phi^\mu}
+{1 \over 2} \PV{\cal L\ \ }{\ODOTS{a}{\alpha\beta}}
\PV{a^{\alpha\beta}}{\phi^\mu} = a_\mu^{\ \alpha}\,p_\alpha 
+ S_{\mu\beta}\,x^\beta\ . \eqno(32)$$
Taking the derivative of (32) with respect to the affine parameter
yields the familiar Weysenhoff condition, that the spin is pure
spacelike in the rest frame of the particle,
$$p_\mu\, S^{\mu\nu} = 0\ . \eqno(33)$$
This is quite significant, because usually (33) is imposed at the onset
by fiat, while we have provided an actual derivation based on the new
automorphism symmetry of the Lagrangian!

\subsection*{C.  Polyrotational Quasi-Coordinates and Quantization}
We propose the body frame
coordinates may be rotated by a generalized geometamorphic 
transformation, a combination of (8) and (21b) which leaves the
Lagrangian (27) invariant.  This provides clues as to the nature
of the derivative with respect to the bivector coordinate.  With
this we can define a new `spin operator' and  generalized
quantum wave equations based on (15).

\subsubsection*{1.  Polydimensionally Rotating Coordinate Frames}
We propose a polydimensional generalization of Section III.A.3,
invoking our principles of polydimensional isotropy and relative
dimensionalism discussed in Section II above.
The poly-rotation operator is,
$$\RRR(\kappa)\equiv \exp\bigl[-(1/4)\,
\EBH{\mu\nu}\Theta^{\mu\nu}(\kappa)
-(1/2)\,\EBH{\alpha}\Phi^\alpha(\kappa)\,\bigr] \ . \eqno(34a)$$
We propose that the {\it polyvelocity} is defined in analogy to
\EQN{22},
$$\ODOT{\bf \cal Q}\equiv {\cal M}/m_0 \equiv
 -2 \lambda\, \RRR^{-1} \ODOT{\RRR}
\ \equiv\ \ODOTS{x}{\mu}\EBH{\mu} + (2\lambda)^{-1}
\ODOTS{a}{\mu\nu}\EBH{\mu\nu} \ , \eqno(34b)$$
where $\{x^\mu, a^{\alpha\beta}\}$ must therefore be anholonomic
quasi-coordinates (as opposed to the holonomic coordinates
$\lambda\Phi^\mu$ and $\lambda^2\Theta^{\alpha\beta}$
respectively),
such that their derivatives with respect to $d\kappa$
yield the vector and bivector velocities.  In analogy to the
rotational development in Section III.A.3 above, the linear velocity
is NOT the (total) derivative of the usual cartesian
coordinate $X^\mu(\kappa)$ for a spinning body.  For example, when
moving in the $y$ direction, an angular acceleration of the spin
along the $z$ axis will introduce an additional ``effective''
velocity in the $x$ direction due to the shift in apparent
relativistic mass center.  All of these interdependent effects are
accounted for in the history-dependent quasi-coordinate $x^\mu$.

It follows that the
tangent basis vectors are given in analogy to \EQN{23b},
$$2\lambda \PD{\RRR}{x^\mu} = -\EB{\mu}\RRR
\ =-\RRR\EBH{\mu}\ ,
\phantom{=2\lambda^2\left[\PD{\;}{x^\alpha},\PD{\;}{x^\beta}\right]
\RRR }
 \eqno(35a)$$
$$2\lambda^2 \PD{\RRR}{a^{\alpha\beta}} = -\EB{\alpha\beta}\RRR
=-\RRR\EBH{\alpha\beta}
=2\lambda^2\left[\PD{\;}{x^\alpha},\PD{\;}{x^\beta}\right]
\RRR \ . \eqno(35b)$$
This implies that the bivector derivative is equivalent to the 
commutator derivative, an idea developed further by Erler\cite{Erler99}
and utilized in Section IV below.  Two other commutators we shall 
find useful are,
$$\left[\PD{\ }{x^\mu}, \PD{\quad}{a^{\alpha\beta}}\right]\RRR
=-\left( g_{\mu\sigma} \over \lambda^2 \right) 
\delta^{\omega\sigma}_{\alpha\beta} \PD{\ }{x^\omega} \RRR\ ,
\quad\quad\quad\quad \eqno(35c)$$
$$\left[\PD{\quad}{a^{\alpha\beta}},\PD{\quad}{a^{\mu\nu}}\right]
\RRR= -\left(g^{\phi\kappa} \over \lambda^2\right)
g_{\sigma\mu}\, g_{\omega\nu}\, 
\delta^{\sigma\omega\theta}_{\alpha\beta\kappa}\,
\PD{\quad}{a^{\theta\phi}}\RRR\ . \eqno(35d)$$
We note in passing that the 4 basis vectors and 6 bivectors of a 4D
space can be reinterpreted as the 10 bivectors of an enveloping 5D
space, with \EQN{34a} as the rotation operator.  In this context
our calculus may be related to that proposed by Blake\cite{Blake}
over the multivector manifold $spin^{+}(4,1)$.

By parallel argument to \EQN{25ab} we obtain the non-commutivity
of the variation and derivative for the polyvelocity,
$$\left(\delta \DDD - \DDD \delta \right) {\bf \cal Q}
= \RRR^{-1} \left(\DDD\RRR\,\delta{\bf \cal Q}
-\delta\RRR\,\DDD{\bf \cal Q} \right)\;=
\;\left[ \DDD{\bf \cal Q} ,\delta{\bf \cal Q}\right]
/(2\lambda^2) \ . \eqno(36)$$
Extracting the vector and bivector portions,
$$\left(\delta d - d \delta \right) x^\alpha =
\left(\delta x^\mu\, da_\mu^{\ \alpha} 
- dx^\mu\, \delta a_\mu^{\ \alpha}  \right)
/\lambda^2\ ,
\ \phantom{+(dx^\mu \delta x^\nu - \delta x^\mu dx^\nu )}
\eqno(37a)$$
$$\left( \delta d - d \delta \right) a^{\mu\nu} = 
\left(da^{\mu\sigma} \delta a_\sigma^{\ \nu} -
\delta a^{\mu\sigma} d a_\sigma^{\ \nu}\right)/\lambda^2
+(dx^\mu \delta x^\nu - \delta x^\mu dx^\nu ) \ . \eqno(37b)$$
Comparing to (20d) and (25b) we see that
$\ODOT{a}\simeq \omega\lambda^2$.
However, we have additional terms which validates that 
variations of the linear and rotational paths are not independent,
hence \EQN{20c} is no longer complete.

\subsubsection*{2.  Quantization}
In classical Hamilton mechanics, functions of motion (n.b. the 
Hamiltonian on which quantum mechanics is based) are parameterized
in terms of the coordinates and their canonical momenta.  The
obvious generalization of the Poisson Bracket for two functions
of polydimensional coordinates would be,
$$\{ F, G \} \equiv \left( \PV{F\ }{x^\alpha} \PV{G\ }{p_\alpha}
- \PV{G\ }{x^\alpha} \PV{F\ }{p_\alpha} \right)+
{\lambda^2 \over 2}
 \left( \PV{F\ }{a^{\alpha\beta}} \PV{G\ }{S_{\alpha\beta}}
- \PV{G\ }{a^{\alpha\beta}}
\PV{F\ }{S_{\alpha\beta}} \right)\ . \eqno(38)$$
There are some potential complications.  From
\EQN{37ab} its not
at all clear that $\delta p^\sigma$ is completely independent of
$\delta x^\mu,\> \delta a^{\alpha\beta} \hbox{ and especially }
\delta S^{\alpha\beta}$.  For today we will sidestep the issues
and assume for brevity that at least the canonical pairs obey
the relations: 
$\{x^\alpha,p_\beta \}=\delta^\alpha_\beta$, and
$\{a^{\mu\nu}, S_{\alpha\beta} \}=
\lambda^2\delta^{\mu\nu}_{\alpha\beta}$.

The {\it Heisenberg quantization rule} is that the commutator of
quantum operators maps to the Poisson bracket of the
corresponding classical quantities,
$$[ \hat{F}, \hat{G}] \mapsto i \hbar \{ F, G \}\ . \eqno(39)$$
It follows that $[ \hat{x}^\nu ,\hat{p}_\mu ] = i \hbar
\delta^\nu_{\ \mu}$ and $[ \hat{a}^{\mu\nu} ,
\hat{S}_{\alpha\beta} ] = i \hbar \lambda^2
\delta^{\mu\nu}_{\alpha\beta}$.  In the coordinate representation
the momenta operators must be,
$$\hat{p}_\mu \equiv -i \hbar \PDE{x^\mu}\ , \quad\quad\quad
\hat{S}_{\mu\nu} \equiv -i\hbar \lambda^2 \PDE{a^{\mu\nu}}
\ . \eqno(40ab)$$
This would imply that one could define a spin angular coordinate
$\theta^{\mu\nu} =\lambda^{-2}\, a^{\mu\nu}$.  We should note that
other authors see potential difficulties with the definition
of angular operators in quantum mechanics\cite{Torre}.

\subsubsection*{3.  (Hand) Wave Equations}
The polydimensional analogy of the Klein-Gordon wave equation
based on the Dixon equation (15) would hence be,
$$\left[ \hat{p}^\mu\,\hat{p}_\mu -{1 \over 2 \lambda^2}
\hat{S}^{\mu\nu}\,\hat{S}_{\mu\nu}\right]
\psi=-\hbar^2\left[\PDE{x^\mu}\PDE{x_\mu}-
{\lambda^2 \over 2} \PDE{a^{\mu\nu}}\PDE{a_{\mu\nu}} 
 \right] \psi =(m_0c)^2 \psi  \ , \eqno(41a)$$
where the wavefunction $\psi(x^\mu,a^{\alpha\beta})$
depends upon the vector position \underbar{and}
bivector spin coordinates.  If the system is in an eigenstate
of total spin, then \EQN{41a} simply reduces to the standard
Klein-Gordon equation with spin-enhanced mass given by \EQN{16a}.

One might expect that the generalization of the Dirac equation would
simply involve factoring (15) with the polymomenta (12b) into
the linear form: $\hat{\cal M}\Psi = m_0c\Psi$.  Its
not quite that simple because we know the components of the standard
spin operator $\hat{\bf S}=(1/2) \hat{S}^{\alpha\beta}\EB{\alpha\beta}$
do not commute.  Consistent with (35d), we have the standard
relations\cite{Khriplovich},
$$\hat{\bf S}\hat{\bf S}= -(1/2)\hat{S}^{\mu\nu} \hat{S}_{\mu\nu} +
\EB{\mu\nu}\EB{\alpha\beta}
[ \hat{S}^{\mu\nu} , \hat{S}^{\alpha\beta} ]
=-(1/2)\hat{S}_{\mu\nu}\hat{S}^{\mu\nu}
-2i\hbar\lambda^{-2} {\bf S}\ . \eqno(42a)$$
Equation (35b) implies that the components of the momenta no longer
commute:
$[\hat{p}_\mu, \hat{p}_\nu]=-i\hbar \hat{S}_{\mu\nu}/\lambda^2$,
such that the square of the momentum vector $\hat{\bf p}=\hat{p}^\mu
\EB{\mu}$,
$$\hat{\bf p}\hat{\bf p}= \hat{p}^\mu \hat{p}_\mu +
\HALF \EB{\mu\nu}\left[ \hat{p}^\mu , \hat{p}^\mu \right]
\ =\ \hat{p}^\mu \hat{p}_\mu -
i\hbar\lambda^{-2}  \hat{\bf S}\ . \eqno(42b)$$
Equation (35c) implies that the spin and momenta operators
do not commute,
$$[ \hat{p}^\mu , \hat{S}_{\alpha\beta} ]
=-i\hbar \delta^{\mu\sigma}_{\alpha\beta}\>\hat{p}_\sigma \ ,
\eqno(42c)$$
$$\{ \hat{\bf p},\hat{\bf S} \} = 
2\hat{\bf p}\wedge\hat{\bf S} -3i\hbar \lambda^{-1} 
\hat{\bf p}\ . \eqno(42d)$$
Putting \EQN{42abd} together, the polymomenta operator (12b) obeys,
\def\MH{{\cal \widehat{M}}}
$$\MH\MH= \left(\hat{p}^\mu\,\hat{p}_\mu
- {1 \over 2\lambda^2}
\hat{S}^{\mu\nu}\,\hat{S}_{\mu\nu}\right) - {3i \hbar \over \lambda}
\MH + {2 \over \lambda} 
\hat{\bf p}\wedge\hat{\bf S} \ . \eqno(43)$$
Substituting, we can rewrite \EQN{41a} as,
$$\left[\MH\left(\MH+ {3i\hbar \over \lambda}\right)
-{2 \over \lambda}\hat{\bf p}\wedge\hat{\bf S} -(m_0c)^2
\right] \psi =0 \ . \eqno(41b)$$
If we presume an idempotent structure on the wavefunction, the
trivector term can be replaced by an eigenvalue,
$${2 \over \lambda} \hat{\bf p}\wedge\hat{\bf S}
\left[\left(1 \pm {i \hat{\bf p}\wedge\hat{\bf S} \over
mcS } \right) \psi \right]\ \Rightarrow
\ \pm \left({2 i mS \over \lambda}\right) \psi \ , \eqno(44)$$
where $S \equiv \parallel{\bf S}\parallel=(1/2)S^{\mu\nu}S_{\mu\nu}$.
Thus \EQN{41b} can now be factored into a polydimensional monogenic
Dirac equation with complex mass roots $N_\pm$,
$$\left[\MH + N_\pm \right] \Psi=
\left[-i\hbar\left(\EBU{\mu}\PDE{x^\mu} +
{\lambda \over 2} \EBU{\mu\nu} \PDE{a^{\mu\nu}}
\right) + N_\pm \right] \Psi=0 \ . \eqno(45a)$$
$$\Psi \equiv \left[\MH - N_\mp \right] \psi \ . \eqno(45b)$$
Solving the quadratic equation, we can get one of the roots to be the
bare mass if we impose a constraint on the magnitude of spin,
$$N_+=m_0 c\ , \quad\quad N_- = m_0 c - 3i\hbar/\lambda \ ,\eqno(45cd)$$
$$S \equiv (1/2) S^{\mu\nu}\,S_{\mu\nu}\, = (3/2)\hbar (m_0/m)
\ . \eqno(45e)$$
Invoking \EQN{16a}, we can thus get a relationship between the
fundamental constants $m_0, \lambda$ and the magnitude of the spin $S$.
In the limit of $m_0c\lambda >>\hbar$ one recovers the standard
`half integer spin' magnitude equation: $S^2=(3/2)\hbar^2$.

\section*{IV.  General Poly-Covariance}
In curved space, particles will now deviate from standard geodesics
due to contributions from derivatives of the basis vectors with
respect to the new bivector coordinate.  Further, there are additional 
contributions to the non-commutivity of the variation and derivative
due to torsion and curvature.  This leads to a new derivation of the
Papapetrou equations\cite{Papa} describing the motion of 
spinning particles in curved space.  Finally we propose a principle
of {\bf Metamorphic Covariance:} {\it that the laws of physics should
be form invariant under local automorphism transformations which 
reshuffle the geometry\/}.

\subsection*{A.  Covariant Derivatives in the Clifford Manifold}
The total derivative of a basis vector with respect to the new
affine parameter $d\kappa$ must by the chain rule contain a
derivative with respect to the bivector coordinate,
$$\ODOT{\bf e}_\mu \equiv {d \EB{\mu} \over d \kappa} =
\ODOTS{x}{\sigma} \PD{\EB{\mu}\ }{x^\sigma} + {1 \over 2}
\ODOTS{a}{\alpha\beta} \PD{\EB{\mu}}{a^{\alpha\beta}}\ . \eqno(46a)$$
Our ans\"atze, consistent with (35b), is
that the bivector derivative obeys\cite{Erler99, Pezz97},
$$\PD{\EB{\mu}\ }{a^{\alpha\beta}}\equiv
\left( \left[ \partial_\alpha,\partial_\beta \right]
-\tau^\sigma_{\alpha\beta} \partial_\sigma \right) \EB{\mu}
=\left( R_{\alpha\beta\mu}^{\ \ \ \nu} - \tau^\sigma_{\alpha\beta}
\Gamma_{\sigma\mu}^\nu \right)\EB{\nu}\ , \eqno(46b)$$
where $\tau^\sigma_{\alpha\beta}$ is the torsion,
$\Gamma_{\sigma\mu}^\nu $ the Cartan connection and 
$R_{\alpha\beta\mu}^{\ \ \ \ \nu}$ the Cartan curvature.

We can factor out the basis vectors by defining the covariant derivative,
$${\partial \ \ \over \partial x^\mu }
\left(p^\nu \EB{\nu} \right) = \EB{\nu} \COV{\mu} p^\nu \equiv\EB{\nu}
\left( \partial_\mu p^\nu + p^\sigma \Gamma^\nu_{\mu\sigma}
\right)\ , \eqno(47a)$$
$${\partial \ \ \ \over \partial a^{\alpha\beta} }
\left(p^\nu \EB{\nu} \right) = \EB{\nu} 
\left[ \COV{\alpha},\COV{\beta} \right] p^\nu
\equiv \EB{\nu} \left( R_{\alpha\beta\mu}^{\quad\ \nu} p^\mu -
\tau^\sigma_{\alpha\beta} \COV{\sigma} p^\nu \right)\ . \eqno(47b)$$
{From} these definitions it is clear than the covariant derivatives of
the basis vectors $\EBU{\mu}$ and $\EB{\mu}$ vanish as usual.

The parallel transport of the conserved canonical momenta generates
new {\it poly-autoparallels} in the Clifford manifold,
$$0={d \over d\kappa } \left( \EB{\mu} p^\mu\right)=
\EB{\mu}\left(\ODOTS{x}{\sigma}\COV{\sigma} +\HALFM
\ODOTS{a}{\alpha\beta} \left[\COV{\alpha},\COV{\beta} \right]
 \right)p^\mu\ , \eqno(48a)$$
$$0={d \over d\kappa } \left( \EB{\mu\nu}S^{\mu\nu}\right)=
\EB{\mu\nu}\left(\ODOTS{x}{\sigma}\COV{\sigma} +\HALFM
\ODOTS{a}{\alpha\beta} \left[\COV{\alpha},\COV{\beta} \right]
 \right)S^{\mu\nu}\ . \eqno(48b)$$
Substituting (47ab) provides a new derviation of
the Papapetrou equations of motion for spinning particles\cite{Papa}
in contravariant form.  Ours however are more general as they include
torsion and all the higher order terms.  In covariant form,
$$0=\ODOTL{p}{\sigma} -\left(\ODOTS{x}{\alpha}
\Gamma_{\alpha\sigma}^{\ \mu} + \HALFM \ODOTS{a}{\alpha\beta}
R_{\alpha\beta\sigma}^{\,\prime\quad\mu}\right) p_\mu\ ,\eqno(49a)$$
$$0=\ODOTL{S}{\rho\omega} -\delta^{\sigma\nu}_{\rho\omega}
\left(\ODOTS{x}{\alpha}
\Gamma_{\alpha\sigma}^{\ \mu} + \HALFM \ODOTS{a}{\alpha\beta}
R_{\alpha\beta\sigma}^{\,\prime\quad\mu}
\right) S_{\mu\nu} \ ,\eqno(49b)$$
$$R_{\alpha\beta\nu}^{\,\prime\quad\mu}\equiv
R_{\alpha\beta\nu}^{\quad\mu}-\tau_{\alpha\beta}^\sigma
\Gamma_{\sigma\nu}^{\ \mu}\ . \eqno(49c) $$

\subsection*{B.  An-Holonomic Mechanics}
It has been a long-standing unsolved problem to derive the Papapetrou
equations from a simple Lagrangian.
We succeed where so many others have failed because of our definition
of the new affine parameter, the form of the Lagrangian (27)
and by noting that the introduction of the
bivector coordinate has made the system an-holonomic.  Consider the
variation of the Lagrangian,
$$\delta{\cal L} = \PV{\cal L\ }{x^\alpha}\>\delta x^\alpha +
\PV{\cal L\ \ }{\ODOTS{x}{\alpha}}\>\VARO{x}{\alpha}
+\HALF \PV{\cal L\ \ }{a^{\alpha\beta}}\>\delta a^{\alpha\beta}+
\HALF \PV{\cal L\ \ }{\ODOTS{a}{\alpha\beta}}
\>\VARO{a}{\alpha\beta}\ . \eqno(50)$$
As in (20b), we must integrate the spin-velocity term by parts,
$$\PV{\cal L\ }{\ODOTS{a}{\alpha\beta}}\VARO{a}{\alpha\beta}=
S_{\alpha\beta}\,\VARO{a}{\alpha\beta}=
{d \over d\kappa} \left(S_{\alpha\beta}\,\delta a^{\alpha\beta} \right) -
\ODOTL{S}{\alpha\beta}\,\delta{a}^{\alpha\beta}
+S_{\alpha\beta} \left(\VARO{a}{\alpha\beta} - {d \over d\kappa}
\delta a^{\alpha\beta} \right). \eqno(51)$$
However, the generalized equation of motion \EQN{20c} is
incomplete because in general there will be an interdependence
between the vector and bivector variations in curved space.
Certainly we saw this feature appear before in \EQN{37ab} for
the polyrotating coordinate system.  The difficulty is how to derive
the new contributions due to torsion and curvature.  

Note while (25b) states that the variation of the angular
velocity is not the velocity of the angular variation, {\it for
the components}, one can easily show from (24ab) and (25ab) that
${\bf d}(\delta \omega^{\mu\nu} \EB{\mu\nu})=
\delta(d\omega^{\mu\nu} \EB{\mu\nu})$.  We therefore argue that
in curved space the same idea holds,
$$\delta \left( \ODOTS{x}{\mu}\>\EB{\mu} \right) =
{d \over d\kappa} \left( \delta x^\mu\>\EB{\mu} \right)\ , \eqno(52a)$$
$$\delta \left( \ODOTS{a}{\alpha\beta}\>\EB{\alpha}\wedge\EB{\beta}
\right) = {d \over d\kappa} \left( \delta a^{\alpha\beta} 
\>\EB{\alpha}\wedge\EB{\beta} \right)\ . \eqno(52b)$$
Performing the variations and derivatives in the above
equations and rearranging terms [and ignoring the contribution
of \EQN{37ab}],
$$\left(\VARO{x}{\mu} - {d \delta x^\mu \over d\kappa\ } \right) =
\delta x^\alpha \ODOTS{x}{\beta} \tau_{\alpha\beta}^\sigma
+ \HALF \left(\delta x^\alpha \ODOTS{a}{\mu\nu} -
\ODOTS{x}{\alpha}\delta a^{\mu\nu} \right)
 R_{\mu\nu\alpha}^{\,\prime\quad\sigma}\ , \eqno(53a)$$
$$\left(\VARO{a}{\mu\nu} - {d \delta a^{\mu\nu}\over d\kappa\ }\right) =
\delta^{\lambda\sigma}_{\omega\nu} \left[ \Gamma^\omega_{\alpha\mu}
\left(\ODOTS{x}{\alpha}\delta a^{\mu\nu} - \ODOTS{a}{\mu\nu}
\delta x^\alpha \right) + {1 \over 4}
R_{\alpha\beta\mu}^{\prime\ \ \,\omega} \left(\ODOTS{a}{\mu\nu}
\delta a^{\alpha\beta} - \ODOTS{a}{\alpha\beta}
\delta a^{\mu\nu} \right) \right] \ . \eqno(53b)$$
The first term on the right of (53a) involving the torsion follows
Kleinert\cite{Kleinert}, the rest are new. Substituting (53a) into (20b)
and (52b) into (51), and finally into (50), separating out terms
proportional to $\delta x^\mu$ and $\delta a^{\mu\nu}$ respectively,
we obtain polydimensionally generalized Euler-Lagrange equations,
$$\PV{\cal L\ }{x^\mu} - \ODOTL{p}{\mu} + p_\lambda
\,\ODOTS{x}{\alpha}\,\tau^\lambda_{\alpha\mu} + \left(
\HALFM p_\lambda\>R_{\alpha\beta\mu}^{\prime\ \ \ \lambda}
+S_{\omega\beta}\>\Gamma^\omega_{\mu\alpha} \right)
\>\ODOTS{a}{\alpha\beta} = 0\ , \eqno(54a)$$
$$\PV{\cal L\ }{a^{\mu\nu}}-\ODOTL{S}{\mu\nu} + p_\lambda
\ODOTS{x}{\sigma} R_{\mu\nu\sigma}^{\prime\ \ \ \lambda}
+\HALFM \left(S_{\omega\beta}R_{\mu\nu\alpha}^{\prime\ \ \ \omega}
-S_{\omega\nu}R_{\alpha\beta\mu}^{\prime\ \ \ \omega}
\right)\ODOTS{a}{\alpha\beta} = 0 \ . \eqno(54b)$$
The first two terms of \EQN{54a} are standard, the third term appears
in Kleinert\cite{Kleinert}, the rest of (54a) and all of (54b) are new.
Explicitly performing the derivative on the Lagrangian in (54a) we
recover the Papapetrou equation (49a).  To get the
spin equation (49b) from (54b) we must introduce
a generalization of \EQN{46b} for bivector variations,
$$\PV{\cal L}{a^{\mu\nu}}\equiv \left[ \PVE{x^\mu},\PVE{x^\nu}
\right] \LLL - \tau^\sigma_{\mu\nu}\PV{\LLL}{x^\sigma}
=\PV{\LLL}{g_{\alpha\beta}}\left( R_{\mu\nu\alpha\beta}^\prime
+R_{\mu\nu\beta\alpha}^\prime \right) \ . \eqno(54c)$$
If there is no torsion, the $R^\prime$ reduces to the Riemann curvature,
which is antisymmetric in the last two indices, hence \EQN{54c} vanishes.

\subsection*{C.  Metamorphic Covariance}
Our Lagrangian (27) is invariant under {\it local} automorphism
transformations, where in general the $\Phi^\mu$ of (34)
can be position dependent upon a path-dependent (history dependent)
integral of a gauge field $B^\nu_{\ \mu}$,
$$\Phi^\nu(x^\alpha) = \int^{x^\alpha} 
B^\nu_{\ \mu}(y^\sigma)\>dy^\mu \ . \eqno(55)$$
This would imply that the connection of a basis vector would become
{\it geometamorphic}\/\cite{Pezz98}, e.g. under parallel transport
{\it a vector will metamorph into a plane}.  We have previously
proposed\cite{Pezz97} such a``metamorphic
Clifford connection'' of the form,
$${\cal D}\EB{\mu}\equiv dx^\alpha \left(\Gamma_{\alpha\mu}^\nu
+\HALF \TENSORL{\Xi}{\alpha\mu}{\nu\sigma} \EB{\nu\sigma}\right)
+\HALF da^{\alpha\beta}\left(\TENSORL{R}{\alpha\beta\mu}{\nu}
\EB{\nu} + \HALF\TENSORL{\Omega}{\alpha\beta\mu}{\nu\sigma}
\EB{\nu\sigma}\right) \>, \eqno(56a)$$
where $\TENSORL{\Xi}{\alpha\mu}{\nu\sigma} \simeq
\TENSORU{B}{\omega}{\alpha} \delta^{\nu\sigma}_{\omega\mu}$, and
the curvature $\TENSORL{R}{\alpha\beta\mu}{\nu}$ now has contributions
from derivatives on both $\Gamma^\nu_{\alpha\mu}$ and
$\TENSORL{\Xi}{\alpha\mu}{\nu\sigma}$. 
This means that equations (48ab) are no longer valid because each only
contains a \underbar{single} dimensional piece.  We are forced to
implement {\it dimensional democracy}, and write our equations only with
{\it polyvectors}.  Further one finds that the Leibniz rule does not
hold over the wedge (or dot) product, although it is valid for the
Clifford (direct) product\cite{Pezz97}.  Hence the metamorphic
connection on the bivector would be computed,
$${\cal D}(\EB{\mu}\wedge\EB{\nu})=\HALF[({\cal D}\EB{\mu}),\EB{\nu}]
+\HALF[\EB{\mu},({\cal D}\EB{\nu})]
\not=({\cal D}\EB{\mu})\wedge\EB{\nu}
+\EB{\mu}\wedge({\cal D}\EB{\nu})\>.\eqno(56b)$$

With these generalizations, reworking Section IV.A, one can get a
poly-covariant generalization\cite{Pezz97}
of the Papapetrou equation (49a),
$$\ODOTS{p}{\mu}+p^\nu\left(\ODOTS{x}{\beta}\Gamma_{\beta\nu}^\mu
+\HALF \ODOTS{a}{\alpha\beta} \TENSORL{R}{\alpha\beta\nu}{\mu}\right)
+\HALF \TENSORU{S}{\omega}{\sigma}\left(\ODOTS{x}{\alpha}
\TENSORL{\Xi}{\alpha\omega}{\mu\sigma}+\HALF\ODOTS{a}{\alpha\beta}
\TENSORL{\Omega}{\alpha\beta\omega}{\mu\sigma}\right)=0. \eqno(57)$$
To derive \EQN{57} from a Lagrangian requires us to make the theory
fully covariant under {\it general} polydimensional coordinate
transformations.  This will cause the quadratic form (26b) to 
acquire cross terms such that the Lagrangian would generalize to,
$$\LLL=m_0c\sqrt{\ODOTS{x}{\alpha}g_{\alpha\beta}\ODOTS{x}{\beta}
+\HALF \ODOTS{x}{\alpha}h_{\alpha\mu\nu}\ODOTS{a}{\mu\nu}
+{1 \over 4 m_0^2 \lambda^4}\ODOTS{a}{\alpha\beta}
\>{\cal I}_{\alpha\beta\mu\nu}\>\ODOTS{a}{\mu\nu}} \ , \eqno(58)$$
where ${\cal I}$ plays the role of the relativistic moment of 
inertia tensor.  This and the interdimensional metric
$h_{\alpha\mu\nu}$ will
cause the linear momenta not to be parallel to the velocity and
spin momenta not parallel to bivector (angular) velocity.

Equation (56a) is the classical analog to the spin covariant covariant
derivative for the Dirac equation derived from generalized automorphism
transformations of the Dirac algebra by Crawford\cite{Crawford},
$$\COV{\mu}=\partial_\mu + i\left( eA_\mu + \gamma^5 a_\mu \right) 
+\gamma_\nu \left(\HALF B^\nu_{\ \mu} + \gamma^5\,i\,b^\nu_{\ \mu}
\right) +\HALF \gamma_{\alpha\beta}\,C^{\alpha\beta}_{\ \mu}
\ , \eqno(59a)$$
$$\left(-i\hbar \gamma^\mu \COV{\mu} 
-m c \right) \psi =0\ . \eqno(59b)$$
The gauge field $B^\mu_{\ \sigma}$ is the same as in \EQN{55}.  In the 
Dirac equation (59b), the usual momentum operator \EQN{40a} has been
replaced by the gauge-covariant
derivative:  $p_\mu \rightarrow -i\hbar \COV{\mu}$.  To get the
interacting form of the polydimensional Dirac equation (45a) we
have suggested\cite{Pezz9902} that one need only additionally
replace the spin operator (40b) with the commutator 
covariant derivative:
$S_{\mu\nu} \rightarrow -i\hbar \lambda^2
\left[\COV{\mu}, \COV{\nu}\right]$,
$$\left(-i\hbar \gamma^\mu \COV{\mu} -i \hbar {\lambda \over 2}
\gamma^{\alpha\beta}\left[\COV{\alpha},\COV{\beta} \right]
-m_0 c \right) \Psi =0\ . \eqno(60)$$
Certainly one could include higher order triple commutator 
derivatives.  In flat space with all but the electromagnetic gauge
field $A_\mu$ suppresed in (59a), the bivector (commutator)
derivative will introduce an anomalous magnetic moment interaction
which provides a possible interpretation of the constant 
$\lambda$.  It remains to be shown that an application of Ehrenfest's
theorem to \EQN{60} can recover the equation of motion (57), in 
anology to the derivation of the Papapetrou equation (49a) from (59b)
by Crawford\cite{Crawford2}.

\section*{V.  Summary}
In introducing {\it Dimensional Democracy}\/ we have given the bivector
a coordinate and shown its utility in the treatment of the classical
spinning particle problem.  This system is invariant under
``polydimensional'' transformations which reshuffle geometry such
that `what is a vector' is {\it Dimensionally Relative}\/ to the
observer's frame.  A fundamentally new action principle has been
introduced which is {\it Polydimensionally Isotropic}.  Generalized
{\it Metamorphic Covariant} equations of motion and quantum wave
equations have been derived which include curvature, torsion and spin.
Most important, the principles proposed have potential broad math and
physics applications beyond the examples in this paper.

\SWITCH{
\section*{VI. Acknowledgments}
Financial support for attending the conference was provided by
Santa Clara University, Department of Physics.
Particular thanks to the {\it Bazaar Caf\'e} of San Francisco
for providing the creative environment in which the majority of this
work was completed, and T. Erler (UC Santa Barbara) for numerous
discussions during the development of most of the basic ideas.
We appreciate J. Crawford's (Penn. State Univ.) help in showing that
some of the results could be rewritten using standard covariant
derivatives.
}{}

\end{document}